\documentclass[paper]{JHEP}
\usepackage{epsfig}
\usepackage{amssymb}
\def\beq{\begin{equation}}
\def\eeq{\end{equation}}

\def\beqn{\begin{eqnarray}}
\def\eeqn{\end{eqnarray}}

\def\HW{{\small HERWIG}}
\def\NLO{{\small NLO}}
\def\MC{{\small MC}}

\def\yes{$\checkmark$}
\def\no{$\times$}
\newcommand\sss{\scriptscriptstyle\rm}

\newcommand\MCatNLO{{\rm MC}@{\rm NLO}}
\newcommand\code{\tt}
\newcommand\variable{\tt}
\newcommand\MSbar{{\overline {\rm MS}}}
\newcommand\sinthW{\sin\theta_{\sss W}}
\newcommand\sinsqthW{\sin^2\theta_{\sss W}}
\newcommand\sinfthW{\sin^4\theta_{\sss W}}
\preprint{
 CERN--PH--TH/2006--012\hfill\\
 Cavendish--HEP--06/03\hfill\\
 GEF--TH--1/2006}
\title{\boldmath The MC@NLO 3.2 Event Generator%
\footnote{Work supported in part by the UK Particle Physics and
Astronomy Research Council and by the EU Fourth Framework Programme
`Training and Mobility of Researchers', Network `Quantum Chromodynamics
and the Deep Structure of Elementary Particles',
contract FMRX-CT98-0194 (DG 12 - MIHT).}}
\author{Stefano Frixione\\
  INFN, Sezione di Genova,
  Via Dodecaneso 33, 16146 Genova, Italy, and\\
  PH Department, TH Unit, CERN, 1211 Geneva 23, Switzerland\\
  E-mail: \email{Stefano.Frixione@cern.ch}}
\author{Bryan R.\ Webber\\
  Cavendish Laboratory, 
  J.J. Thomson Avenue, Cambridge CB3 0HE, U.K.\\
  E-mail: \email{webber@hep.phy.cam.ac.uk}}
\abstract{
This is the user's manual of {$\MCatNLO$} 3.2. This package is a 
practical implementation, based upon the HERWIG event generator,
of the $\MCatNLO$ formalism, which allows one to incorporate NLO QCD 
matrix elements consistently into a parton shower framework.
Processes available in this version include the hadroproduction of
single vector and Higgs bosons, vector boson pairs, heavy
quark pairs, single top, lepton pairs, and Higgs bosons in association 
with a $W$ or $Z$. Spin correlations in decays are included for all 
processes except $t\bar{t}$, single-$t$, $ZZ$, and $WZ$ production.
This document is self-contained, but we emphasise the main 
differences with respect to previous versions.
}
\begin{document}

\section{Generalities}
In this documentation file, we briefly describe how to run the 
$\MCatNLO$ package, implemented according to the formalism introduced in 
ref.~\cite{Frixione:2002ik}. When using $\MCatNLO$, please cite
refs.~\cite{Frixione:2002ik,Frixione:2003ei}; if single-top events
are generated, please also cite ref.~\cite{Frixione:2005vw}.
The production processes now available are listed in table~\ref{tab:proc}. 
The process codes {\variable IPROC} will be explained below. 
$H_{1,2}$ represent hadrons (in practice, $p$ or $\bar p$).
The treatment of (undecayed) vector boson pair production within $\MCatNLO$ 
has been described in ref.~\cite{Frixione:2002ik}, that of heavy quark pair 
production in ref.~\cite{Frixione:2003ei}. The NLO matrix elements 
for these processes have been taken from 
refs.~\cite{Mele:1990bq,Frixione:1992pj,Frixione:1993yp,Mangano:1991jk}.
The information given in refs.~\cite{Frixione:2002ik,Frixione:2003ei}
allows the implementation in MC@NLO of any production process, provided
that the formalism of refs.~\cite{Frixione:1995ms,Frixione:1997np} is
used for the computation of cross sections to NLO accuracy. For specific 
information on single-top production, see ref.~\cite{Frixione:2005vw}.
The matrix elements for Standard Model Higgs, single vector boson, 
lepton pair, associated Higgs, and single-top production have been taken from 
refs.~\cite{Dawson:1990zj,Djouadi:1991tk}, ref.~\cite{Altarelli:1979ub}, 
ref.~\cite{Aurenche:1980tp}, ref.~\cite{Oleari:2005inprep}, and
ref.~\cite{Harris:2002md} respectively.
\begin{table}[htb]
\begin{center}
\begin{tabular}{|c|c|c|c|c|l|}\hline
{\variable IPROC} & {\variable IV} & {\variable IL}$_1$ & {\variable IL}$_2$ & 
 Spin & Process \\\hline
 --1350--{\variable IL} & & & &\yes &
 $H_1 H_2\to (Z/\gamma^*\to) l_{\rm IL}\bar{l}_{\rm IL}+X$\\\hline
 --1360--{\variable IL} & & & &\yes &
 $H_1 H_2\to (Z\to) l_{\rm IL}\bar{l}_{\rm IL}+X$\\\hline
 --1370--{\variable IL} & & & &\yes &
 $H_1 H_2\to (\gamma^*\to) l_{\rm IL}\bar{l}_{\rm IL}+X$\\\hline
 --1460--{\variable IL} & & & &\yes &
 $H_1 H_2\to (W^+\to) l_{\rm IL}^+\nu_{\rm IL}+X$\\\hline
 --1470--{\variable IL} & & & &\yes &
 $H_1 H_2\to (W^-\to) l_{\rm IL}^-\bar{\nu}_{\rm IL}+X$\\\hline
 --1396 & & & &\no &
 $H_1 H_2\to \gamma^*(\to \sum_i f_i\bar{f}_i)+X$\\\hline
 --1397 & & & &\no &
 $H_1 H_2\to Z^0+X$\\\hline
 --1497 & & & &\no &
 $H_1 H_2\to W^+ +X$\\\hline
 --1498 & & & &\no &
 $H_1 H_2\to W^- +X$\\\hline
 --1600--{\variable ID} & & & & &
 $H_1 H_2\to H^0+X$\\\hline
 --1705 & & & & &
 $H_1 H_2\to b\bar{b}+X$\\\hline
 --1706 & & & & \no &
 $H_1 H_2\to t\bar{t}+X$\\\hline
 --2000--{\variable IC} & & & & \no &
 $H_1 H_2\to t/\bar{t}+X$\\\hline
 --2001--{\variable IC} & & & & \no &
 $H_1 H_2\to \bar{t}+X$\\\hline
 --2004--{\variable IC} & & & & \no &
 $H_1 H_2\to t+X$\\\hline
 --2600--{\variable ID} & 1 & 7 & &\no &
 $H_1 H_2\to H^0 W^+ +X$\\\hline
 --2600--{\variable ID} & 1 & $i$ & &\yes &
 $H_1 H_2\to H^0 (W^+\to)l_i^+\nu_i +X$\\\hline
 --2600--{\variable ID} & -1 & 7 & &\no &
 $H_1 H_2\to H^0 W^- +X$\\\hline
 --2600--{\variable ID} & -1 & $i$ & &\yes &
 $H_1 H_2\to H^0 (W^-\to)l_i^-\bar{\nu}_i +X$\\\hline
 --2700--{\variable ID} & 0 & 7 & &\no &
 $H_1 H_2\to H^0 Z +X$\\\hline
 --2700--{\variable ID} & 0 & $i$ & &\yes &
 $H_1 H_2\to H^0 (Z\to)l_i\bar{l}_i +X$\\\hline
 --2850 & & 7 & 7 & \no &
 $H_1 H_2\to W^+W^-+X$\\\hline
 --2850 & & $i$ & $j$ & \yes &
 $H_1 H_2\to (W^+\to)l_i^+\nu_i (W^-\to)l_j^-\bar{\nu}_j +X$\\\hline
 --2860 & & 7 & 7 & \no &
 $H_1 H_2\to Z^0Z^0+X$\\\hline
 --2870 & & 7 & 7 & \no &
 $H_1 H_2\to W^+Z^0+X$\\\hline
 --2880 & & 7 & 7 & \no &
 $H_1 H_2\to W^-Z^0+X$\\\hline
\end{tabular}
\end{center}
\caption{\label{tab:proc} 
Processes implemented in $\MCatNLO~3.2$. $H^0$ denotes the Standard Model
Higgs boson and the value of {\variable ID} controls its decay, as described
in the \HW\ manual and below. The values of {\variable IV}, {\variable IL},
{\variable IL}$_1$, and {\variable IL}$_2$ control the identities of vector
bosons and leptons, as described below. In single-$t$ production, the
value of {\variable IC} controls the production processes, as described
below. {\variable IPROC}--10000 generates the same processes as 
{\variable IPROC}, but eliminates the underlying event. A void entry 
indicates that the corresponding variable is unused. The `Spin' column 
indicates whether spin correlations in vector boson or top decays are 
included (\yes), neglected (\no) or absent (void entry). Spin correlations 
in Higgs decays are included by \HW\ (e.g. in 
$H^0\to W^+W^-\to l^+\nu l^-\bar{\nu}$).  }
\end{table}

This documentation refers to $\MCatNLO$ version 3.2 (previous versions
1.0, 2.0, 2.2, 2.3, and 3.1 are described in refs.~\cite{Frixione:2002bd,
Frixione:2003ep,Frixione:2003vm,Frixione:2004wy,Frixione:2005gz}
respectively). The new process implemented since version 3.1 is 
single-top production; furthermore, the use of parton density
library LHAPDF~\cite{Giele:2002hx,Whalley:2005nh} is now supported.
For precise details of version changes, see 
app.~\ref{app:newver}-\ref{app:newvere}.

\subsection{Mode of operation}
In the case of standard MC, a hard kinematic configuration is
generated on a event-by-event basis, and it is subsequently showered 
and hadronized. In the case of $\MCatNLO$, all of the hard kinematic
configurations are generated in advance, and stored in a file 
(which we call {\em event file} -- see sect.~\ref{sec:evfile}); 
the event file is then read by \HW, which showers and hadronizes each 
hard configuration. Since version 2.0, the events are 
handled by the ``Les Houches'' generic user process 
interface~\cite{Boos:2001cv} (see ref.~\cite{Frixione:2003ei} for 
more details). Therefore, in $\MCatNLO$ the reading of a 
hard configuration from the event file is equivalent to the generation 
of such a configuration in the standard MC.

The signal to \HW\ that configurations should be read from an event file using
the Les Houches interface is a negative value of the process code {\variable
IPROC}; this accounts for the negative values in table~\ref{tab:proc}. In the
case of heavy quark pair, Higgs, Higgs in association with a $W$ or $Z$, 
and lepton pair (through $Z/\gamma^*$ exchange) production, the codes 
are simply the negative of those for the corresponding standard \HW\ MC
processes. Where possible, this convention will be adopted for additional
$\MCatNLO$ processes.  Consistently with what happens in standard \HW, by
subtracting 10000 from {\variable IPROC} one generates the same processes as
in table~\ref{tab:proc}, but eliminates the underlying event\footnote{The
same effect can be achieved by setting the {\scriptsize HERWIG} parameter
{\variable PRSOF} $=0$.}.
 
Higgs decays are controlled in the same way as in \HW, that is by adding
{\variable -ID} to the process code. The conventions for {\variable ID}
are the same as in \HW, namely {\variable ID} $=1\ldots 6$ for 
$u\bar{u}\ldots t\bar{t}$; $7\ldots 9$ for $e^+e^-\ldots \tau^+\tau^-$;
$10, 11$ for $W^+W^-, ZZ$; and 12 for $\gamma\gamma$. Furthermore,
{\variable ID} $=0$ gives quarks of all flavours, and {\variable ID} $=99$
gives all decays.

Process codes {\variable IPROC}=$-1360-${\variable IL} and 
$-1370-${\variable IL} do not have an analogue in \HW; they are the 
same as $-1350-${\variable IL}, except for the fact that only a $Z$ or a
$\gamma^*$ respectively is exchanged. The value of {\variable IL} determines
the lepton identities, and the same convention as in \HW\ is adopted:
{\variable IL}=$1,\ldots,6$ for $l_{\rm IL}=e,\nu_e,\mu,\nu_\mu,\tau,\nu_\tau$
respectively. At variance with \HW, {\variable IL} cannot be set equal to
zero. Process codes {\variable IPROC}=$-1460-${\variable IL} and
$-1470-${\variable IL} are the analogue of \HW\ $1450+${\variable IL};
in \HW\, either $W^+$ or $W^-$ can be produced, whereas MC@NLO treats
the two vector bosons separately. For these processes, as in \HW,
{\variable IL}=$1,2,3$ for $l_{\rm IL}=e,\mu,\tau$, but again
the choice ${\variable IL}=0$ is not allowed.

The lepton pair processes {\variable IPROC}=$-1350-${\variable IL},
$\ldots$, $-1470-${\variable IL} include spin correlations when
generating the angular distributions of the
produced leptons. However, if spin correlations are not an
issue, the single vector boson production processes
{\variable IPROC}= $-$1396,$-$1397,$-$1497,$-$1498 can be used,
in which case the vector boson decay products are distributed
according to phase space.

There are a number of other differences between the lepton pair
and single vector boson processes.
The latter do not feature the $\gamma$--$Z$ interference terms. Also, their
cross sections are fully inclusive in the final-state fermions resulting 
from $\gamma^*$, $Z$ or $W^\pm$.  The user can still select a definite
decay mode using the variable {\variable MODBOS} (see sect.~\ref{sec:decay}),
but the relevant branching ratio will {\em not} be included automatically by
MC@NLO.  In the case of $\gamma^*$ production, the branching ratios are $C_i
q_i^2/(20/3)$, $q_i$ being the electric charge (in units of the positron
charge) of the fermion $i$ selected through {\variable MODBOS}, and $C_i=1$
for leptons and 3 for quarks. Notice that $20/3=\sum_i C_i q_i^2$, the sum
including all leptons and quarks except the top. Thus, the total rate
predicted by MC@NLO in the case of lepton pair production can also
be recovered by multiplying the corresponding single vector boson total
rate by the relevant branching ratio.

In NLO computations for single-top production, it is customary to
distinguish between three production mechanisms, conventionally
denoted as $s$ channel, $t$ channel, and $Wt$ mode. In version 3.2,
only the former two are implemented. The user can generate them
either separately, by setting {\variable IC}=$10$ and {\variable IC}=$20$ 
for $s$- and $t$-channel production respectively, or together, by 
setting {\variable IC}=$0$. For example, according to table~\ref{tab:proc},
$t$-channel single-$\bar{t}$ events will be generated by entering
{\variable IPROC}=$-2021$.

In the case of vector boson pair production, the process codes are the 
negative of those adopted in $\MCatNLO$ 1.0 (for which the Les Houches 
interface was not yet available), rather than those of standard \HW.

Furthermore, in the case of Higgs production in association with a $W$ or $Z$,
as well as vector boson pair production, the value of {\variable IPROC} alone
is not sufficient to fully determine the process type, and new variables
{\variable IV}, {\variable IL}$_1$, and {\variable IL}$_2$ have been
introduced (see table~\ref{tab:proc}).  The variables {\variable IL}$_1$ and
{\variable IL}$_2$ can take the same values as {\variable IL} relevant to
lepton pair production (notice, however, that in the latter case {\variable
IL} is not an independent variable, and its value is included via {\variable
IPROC}); in addition, {\variable IL}$_\alpha$=7 implies that lepton spin
correlations for the decay products of the corresponding vector boson are not
taken into account, as indicated in table~\ref{tab:proc}.

Apart from the above differences, $\MCatNLO$ and \HW\ {\em behave in exactly
the same way}. Thus, the available user's analysis routines can be 
used in the case of $\MCatNLO$. One should recall, however, that
$\MCatNLO$ always generates some events with negative weights (see
refs.~\cite{Frixione:2002ik,Frixione:2003ei}); therefore, the correct
distributions are obtained by summing weights with their signs (i.e., the
absolute values of the weights must {\em NOT} be used when filling the
histograms).

With such a structure, it is natural to create two separate executables,
which we improperly denote as \NLO\ and \MC. The former has the sole scope
of creating the event file; the latter is just \HW, augmented by
the capability of reading the event file.

\subsection{Package files\label{sec:packfile}}
The package consists of the following files:

\begin{itemize}
\item {\bf Shell utilities}\\
    {\code MCatNLO.Script}\\
    {\code MCatNLO.inputs}\\
    {\code Makefile}

\item {\bf Utility codes}\\
    {\code MEcoupl.inc}\\ 
    {\code alpha.f}\\ 
    {\code dummies.f}\\ 
    {\code linux.f}\\ 
    {\code mcatnlo\_date.f}\\  
    {\code mcatnlo\_hbook.f}\\  
    {\code mcatnlo\_helas2.f}\\  
    {\code mcatnlo\_hwdummy.f}\\ 
    {\code mcatnlo\_int.f}\\
    {\code mcatnlo\_lhauti.f}\\  
    {\code mcatnlo\_libofpdf.f}\\ 
    {\code mcatnlo\_mlmtolha.f}\\  
    {\code mcatnlo\_mlmtopdf.f}\\ 
    {\code mcatnlo\_pdftomlm.f}\\ 
    {\code mcatnlo\_str.f}\\ 
    {\code mcatnlo\_uti.f}\\ 
    {\code mcatnlo\_uxdate.c}\\
    {\code sun.f}\\ 
    {\code trapfpe.c}

\item {\bf General \HW\ routines}\\
    {\code mcatnlo\_hwdriver.f}\\ 
    {\code mcatnlo\_hwlhin.f}

\item {\bf Process-specific codes}\\
    {\code mcatnlo\_hwanbtm.f}\\
    {\code mcatnlo\_hwanhgg.f}\\
    {\code mcatnlo\_hwanllp.f}\\
    {\code mcatnlo\_hwantop.f}\\
    {\code mcatnlo\_hwanstp.f}\\
    {\code mcatnlo\_hwansvb.f}\\
    {\code mcatnlo\_hwanvbp.f}\\
    {\code mcatnlo\_hwanvhg.f}\\
    {\code mcatnlo\_hgmain.f}\\
    {\code mcatnlo\_hgxsec.f}\\
    {\code mcatnlo\_llmain.f}\\
    {\code mcatnlo\_llxsec.f}\\
    {\code mcatnlo\_qqmain.f}\\
    {\code mcatnlo\_qqxsec.f}\\
    {\code mcatnlo\_sbmain.f}\\
    {\code mcatnlo\_sbxsec.f}\\
    {\code mcatnlo\_stmain.f}\\
    {\code mcatnlo\_stxsec.f}\\
    {\code mcatnlo\_vbmain.f}\\
    {\code mcatnlo\_vbxsec.f}\\
    {\code mcatnlo\_vhmain.f}\\
    {\code mcatnlo\_vhxsec.f}\\
    {\code hgscblks.h}\\
    {\code hvqcblks.h}\\
    {\code llpcblks.h}\\
    {\code stpcblks.h}\\
    {\code svbcblks.h}\\
    {\code vhgcblks.h}
\end{itemize}
These files can be downloaded from the web page:\\
$\phantom{aaaaaaaa}$%
{\code http://www.hep.phy.cam.ac.uk/theory/webber/MCatNLO}\\
The files {\code mcatnlo\_hwan{\em xxx}.f}, which appear in the
list of the process-specific codes, are sample \HW\ analysis
routines. They are provided here to give the user a ready-to-run
package, but they should be replaced with appropriate codes according 
to the user's needs.

In addition to the files listed above, the user will need a
version of the \HW\ code
\cite{Marchesini:1992ch,Corcella:2001bw,Corcella:2002jc}.
As stressed in 
ref.~\cite{Frixione:2002ik}, for the $\MCatNLO$ we do not
modify the existing (LL) shower algorithm. However, since $\MCatNLO$
versions 2.0 and higher make use of the Les Houches interface,
first implemented in \HW\ 6.5, the version must be 6.500 or higher.
On most systems, users will need to delete the dummy  subroutines 
{\small UPEVNT}, {\small UPINIT}, {\small PDFSET} and {\small STRUCTM}
from the standard  \HW\ package, to permit linkage of the corresponding
routines from the $\MCatNLO$ package. As a general rule, the user is
strongly advised to use the most recent version of \HW\ (currently
6.510 -- with versions lower than 6.504 problems can be found in attempting 
to specify the decay modes of single vector bosons through the variable
{\variable MODBOS}. Also, crashes in the shower phase have been reported 
when using \HW\ 6.505, and we therefore recommend not to use that
version).

\subsection{Working environment}
We have written a number of shell scripts and a {\code Makefile} (all
listed under {\bf Shell utilities} above) which will simplify the use of
the package considerably. In order to use them, the computing system
must support {\code bash} shell, and {\code gmake}\footnote{For Macs 
running under OSX v10 or higher, {\code make} can be used instead of 
{\code gmake}.}. 
Should they be unavailable on the user's computing system, the compilation 
and running of $\MCatNLO$ requires more detailed instructions; in this case,
we refer the reader to app.~\ref{app:instr}. This appendix will serve also as
a reference for a more advanced use of the package.

\subsection{Source and running directories}
We assume that all the files of the package sit in the same directory,
which we call the {\em source directory}. When creating the executable, 
our shell scripts determine the type of operating system, and create a
subdirectory of the source directory, which we call the {\em running 
directory}, whose name is {\variable Alpha}, {\variable Sun}, {\variable
Linux}, or {\variable Darwin}, depending on the operating system.  
If the operating system is not known by our scripts, the name of the 
working directory is {\variable Run}. The running directory contains all 
the object files and executable files, and in general all the files produced
by the $\MCatNLO$ while running.  It must also contain the relevant grid files
(see sect.~\ref{sec:pdfs}), or links to them, if the library of parton
densities provided with the $\MCatNLO$ package is used.

\section{Prior to running\label{sec:priors}}
Before running the code, the user needs to edit the following files:\\
$\phantom{aaa}${\code mcatnlo\_hwan{\em xxx}.f}\\
$\phantom{aaa}${\code mcatnlo\_hwdriver.f}\\ 
$\phantom{aaa}${\code mcatnlo\_hwlhin.f}\\
We do not assume that the user will adopt the latest release of \HW\ 
(although, as explained above, it must be version 6.500 or higher). For this 
reason, the files {\code mcatnlo\_hwdriver.f} and {\code mcatnlo\_hwlhin.f} 
must be edited, in order to modify the {\code INCLUDE HERWIGXX.INC} command
to correspond to the version of \HW\ the user is going to adopt.
{\code mcatnlo\_hwdriver.f} contains a set of read statements,
which are necessary for the \MC\ to get the input parameters (see
sect.~\ref{sec:running} for the input procedure); these read
statements must not be modified or eliminated. Also, {\code
mcatnlo\_hwdriver.f} calls the \HW\ routines which
perform showering, hadronization, decays (see sect.~\ref{sec:decay} 
for more details on this issue), and so forth; the user can
freely modify this part, as customary in \MC\ runs. Finally, the sample
codes {\code mcatnlo\_hwan{\em xxx}.f} contain analysis-related routines:
these files must be replaced by files which contain the user's analysis 
routines. We point out that, since version 2.0, the {\code Makefile} need not
be edited any longer, since the corresponding operations are now 
performed by setting script variables (see sect.~\ref{sec:scrvar}).

\subsection{Parton densities\label{sec:pdfs}}
Since the knowledge of the parton densities (PDF) is necessary in
order to get the physical cross section, a PDF library must be
linked. The possibility exists to link the (now obsolete) CERNLIB 
PDF library (PDFLIB), or its replacement LHAPDF;
however, we also provide a self-contained PDF library with this package, 
which is faster than PDFLIB, and contains PDF sets released after the 
last and final PDFLIB version (8.04; most of these sets are now included 
in LHAPDF). A complete list of the PDFs available in our PDF library can 
be downloaded from the MC@NLO web page. The user may link one of the three
PDF libraries; all that is necessary is to set the variable {\variable
PDFLIBRARY} (in the file {\code MCatNLO.inputs}) equal to {\variable
THISLIB} if one wants to link to our PDF library, and equal to
{\variable PDFLIB} or to {\variable LHAPDF} if one wants to link 
to PDFLIB or to LHAPDF.  Our PDF library collects 
the original codes, written by the authors of the PDF fits;
as such, for most of the densities it needs to read the files which
contain the grids that initialize the PDFs. These files, which can
be also downloaded from the $\MCatNLO$ web page, must either be copied 
into the running directory, or defined in the running directory as logical
links to the physical files (by using {\code ln -sn}). We stress that if
the user runs MC@NLO with the shell scripts, the logical links will
be created automatically at run time.

As stressed before, consistent inputs must be given to the \NLO\ and
\MC\ codes. However, in ref.~\cite{Frixione:2002ik} we found that the
dependence upon the PDFs used by the MC is rather weak. So one may
want to run the \NLO\ and \MC\ adopting a regular NLL-evolved set in the
former case, and the default \HW\ set in the latter (the advantage is
that this option reduces the amount of running time of the \MC). In
order to do so, the user must set the variable {\variable HERPDF}
equal to {\variable DEFAULT} in the file {\code MCatNLO.inputs};
setting {\variable HERPDF=EXTPDF} will force the \MC\ to use the same
PDF set as the \NLO\ code.

Regardless of the PDFs used in the \MC\ run, users must delete the dummy 
PDFLIB routines {\small PDFSET} and {\small STRUCTM} from \HW, as
explained earlier.

\section{Running\label{sec:running}}
It is straightforward to run the $\MCatNLO$. First, edit\\
$\phantom{aaa}${\code MCatNLO.inputs}\\
and write there all the input parameters (for the complete list 
of the input parameters, see sect.~\ref{sec:scrvar}). As the last
line of the file {\code MCatNLO.inputs}, write\\
$\phantom{aaa}${\code runMCatNLO}\\
Finally, execute {\code MCatNLO.inputs} from the {\code bash} shell.
This procedure will create the \NLO\ and \MC\ executables, and run them
using the inputs given in {\code MCatNLO.inputs}, which guarantees
that the parameters used in the \NLO\ and \MC\ runs are consistent.
Should the user only need to create the executables without running
them, or to run the \NLO\ or the \MC\ only, he/she should replace the
call to {\code runMCatNLO} in the last line of {\code MCatNLO.inputs}
by calls to\\
$\phantom{aaa}${\code compileNLO}\\
$\phantom{aaa}${\code compileMC}\\
$\phantom{aaa}${\code runNLO}\\
$\phantom{aaa}${\code runMC}\\
which have obvious meanings. We point out that the command {\code runMC}
may be used with {\variable IPROC}=1350+{\variable IL}, 1450+{\variable IL}, 
1600+{\variable ID}, 1699, 1705, 1706, 2000--2008, 2600+{\variable ID}, 2699, 
2700+{\variable ID}, 2799 to generate $Z/\gamma^*$, $W^\pm$, Higgs, 
$b\bar{b}$, $t\bar{t}$, single top, $H^0W$ or $H^0Z$ events with 
standard \HW\ (see the \HW\ manual for more details).

We stress that the input parameters are not solely related to
physics (masses, CM energy, and so on); there are a few of them
which control other things, such as the number of events generated.
These must also be set by the user, according to his/her needs:
see sect.~\ref{sec:scrvar}.

Two such variables are {\variable HERWIGVER} and {\variable HWUTI},
which were moved in version 2.0 from the {\code Makefile} to
{\code MCatNLO.inputs}. The former variable must be set equal 
to the object file name of the version of \HW\ currently adopted 
(matching the one whose common blocks are included in the files
mentioned in sect.~\ref{sec:priors}). The variable {\variable HWUTI} 
must be set equal to the list of object files that the user needs in 
the analysis routines.

If the shell scripts are not used to run the codes, the inputs are
given to the \NLO\ or \MC\ codes during an interactive talk-to phase;
the complete sets of inputs for our codes are reported in 
app.~\ref{app:input} for vector boson pair production.

\subsection{Event file\label{sec:evfile}}
The \NLO\ code creates the event file. In order to do so, it goes through
two steps; first it integrates the cross sections (integration step),
and then, using the information gathered in the integration step, 
produces a set of hard events (event generation step). Integration and
event generation are performed with a modified version of the 
{\small SPRING-BASES} package~\cite{Kawabata:1995th}.

We stress that the events stored in the event file just contain the
partons involved in the hard suprocesses. Owing to the modified subtraction
introduced in the $\MCatNLO$ formalism (see ref.~\cite{Frixione:2002ik}) 
they do not correspond to pure NLO configurations, and should not be 
used to plot physical observables. Parton-level observables must be
reconstructed using the fully-showered events.

The event generation step necessarily follows the integration step;
however, for each integration step one can have an arbitrary number of
event generation steps, i.e., an arbitrary number of event files.
This is useful in the case in which the statistics accumulated 
with a given event file is not sufficient.

Suppose the user wants to create an event file; editing {\code
MCatNLO.inputs}, the user sets {\variable BASES=ON}, to enable the
integration step, sets the parameter {\variable NEVENTS} equal to
the number of events wanted on tape, and runs the code; the
information on the integration step (unreadable to the user, but
needed by the code in the event generation step) is written on files
whose name begin with {\variable FPREFIX}, a string the user sets
in {\code MCatNLO.inputs}; these files (which we denotes as {\em data
files}) have extensions {\code .data}. The name of the event file is 
{\variable EVPREFIX.events}, where {\variable EVPREFIX} is again a 
string set by the user.

Now suppose the user wants to create another event file, to increase
the statistics. The user simply sets {\variable BASES=OFF}, since 
the integration step is not necessary any longer (however, the data
files must not be removed: the information
stored there is still used by the \NLO\ code); changes the string
{\variable EVPREFIX} (failure to do so overwrites the existing event
file), while keeping {\variable FPREFIX} at the same value as before;
and changes the value of {\variable RNDEVSEED} (the random number
seed used in the event generation step; failure to do so results in
an event file identical to the previous one); the number {\variable
NEVENTS} generated may or may not be equal to the one chosen in
generating the former event file(s).

We point out that data and event files may be very large. If the user
wants to store them in a scratch area, this can be done by setting the
script variable {\variable SCRTCH} equal to the physical address
of the scratch area (see sect.~\ref{sec:res}).

\subsection{Decays}\label{sec:decay}
$\MCatNLO$ is intended primarily for the study of NLO corrections
to production cross sections and distributions; NLO corrections to
the decays of produced particles are not included. As for spin 
correlations in decays, the situation in version 3.2 is 
summarized in table~\ref{tab:proc}: they are included for all 
processes except $t\bar{t}$, single-$t$, $ZZ$, and $WZ$ 
production\footnote{Non-factorizable spin correlations of 
virtual origin are not included in $W^+W^-$ production.}.
For the latter processes, quantities sensitive to the polarisation of 
produced particles are not given correctly even to leading order.
For such quantities, it may be preferable to use the standard
\HW\ MC, which does include leading-order spin correlations.

Particular decay modes of vector bosons may be forced in $\MCatNLO$
in the same way as in standard \HW, using the {\variable MODBOS}
variables -- see sect.~3.4 of ref.~\cite{Corcella:2001bw}. However,
top decays cannot be forced in this way because the decay is
treated as a three-body process: the $W^\pm$ boson entry in
{\code HEPEVT} is for information only.  Instead, the top
branching ratios can be altered using the {\variable HWMODK}
subroutine -- see sect.~7 of ref.~\cite{Corcella:2001bw}.
This is done separately for the $t$ and $\bar t$. For example,
{\code CALL HWMODK(6,1.D0,100,12,-11,5,0,0)} forces
the decay $t\to \nu_e e^+ b$, while leaving $\bar t$ decays
unaffected.  Note that the order of the decay products is
important for the decay matrix element
({\variable NME} = 100) to be applied correctly.
The relevant statements should be inserted in the \HW\ main program
(corresponding to {\code mcatnlo\_hwdriver.f} in this package)
after the statement {\code CALL HWUINC} and before
the loop over events.  A separate run with
{\code CALL HWMODK(-6,1.D0,100,-12,11,-5,0,0)} should
be performed if one wishes to symmetrize the forcing of
$t$ and $\bar t$ decays, since calls to {\variable HWMODK} from
within the event loop do not produce the desired result.

\subsection{Results\label{sec:res}}
As in the case of standard \HW\, the form of the results will be
determined by the user's analysis routines. However, in addition
to any files written by the user's analysis routines, the
$\MCatNLO$ writes the following files:\\
$\blacklozenge$ 
{\variable FPREFIXNLOinput}: the input file for the \NLO\ executable, 
created according to the set of input parameters defined in 
{\code MCatNLO.inputs} (where the user also sets the string
{\variable FPREFIX}). See table~\ref{tab:NLOi}.\\
$\blacklozenge$ 
{\variable FPREFIXNLO.log}: the log file relevant to the \NLO\ run.\\
$\blacklozenge$ 
{\variable FPREFIXxxx.data}: {\variable xxx} can assume several different 
values. These are the data files created by the \NLO\ code. They can be 
removed only if no further event generation step is foreseen with the
current choice of parameters.\\
$\blacklozenge$ 
{\variable FPREFIXMCinput}: analogous to {\variable FPREFIXNLOinput}, 
but for the \MC\ executable. See table~\ref{tab:MCi}.\\
$\blacklozenge$ 
{\variable FPREFIXMC.log}: analogous to {\variable FPREFIXNLO.log}, but 
for the \MC\ run.\\
$\blacklozenge$ 
{\variable EVPREFIX.events}: the event file, where {\variable EVPREFIX} 
is the string set by the user in {\code MCatNLO.inputs}.\\
$\blacklozenge$ 
{\variable EVPREFIXxxx.events}: {\variable xxx} can assume several different 
values. These files are temporary event files, which are used by the
\NLO\ code, and eventually removed by the shell scripts. They MUST NOT be
removed by the user during the run (the program will crash or give
meaningless results).

By default, all the files produced by the $\MCatNLO$ are written in the
running directory.  However, if the variable {\variable SCRTCH} (to be set in
{\code MCatNLO.inputs}) is {\em not} blank, the data and event files will be
written in the directory whose address is stored in {\variable SCRTCH}
(such a directory is not created by the scripts, and must already exist
at run time).

\section{Script variables\label{sec:scrvar}}
In the following, we list all the variables appearing in 
{\code MCatNLO.inputs}; these can be changed by the user to suit 
his/her needs. This must be done by editing {\code MCatNLO.inputs}.
For fuller details see the comments in {\code MCatNLO.inputs}.
\begin{itemize}
\item[{\variable ECM}] 
 The CM energy of the colliding particles.
\item[{\variable FREN}] 
 The ratio between the renormalization scale, and a reference mass scale.
\item[{\variable FFACT}] 
 As {\variable FREN}, for the factorization scale.
\item[{\variable HVQMASS}] 
 The mass (in GeV) of the top quark, except when {\variable IPROC}=--(1)1705,
 when it is the mass of the bottom quark. In this case, {\variable HVQMASS} 
 must coincide with {\variable BMASS}.
\item[{\variable xMASS}] 
 The mass (in GeV) of the particle {\variable x}, with 
 {\variable x=HGG,W,Z,U,D,S,C,B,G}.
\item[{\variable xWIDTH}] 
 The physical (Breit-Wigner) width (in GeV) of the particle {\variable x}, 
 with {\variable x=HGG,W,Z}.
\item[{\variable IBORNHGG}] 
 Valid entries are 1 and 2.  If set to 1, the exact top mass dependence is
retained {\em at the Born level} in Higgs production.  If set to 2, the
$m_t\to\infty$ limit is used.
\item[{\variable xGAMMAX}] 
 If {\variable xGAMMAX} $>0$, controls the width of the mass range for 
 Higgs ({\variable x=H}) and vector bosons ({\variable x=V1,V2}): the range is 
 ${\variable MASS}\pm({\variable GAMMAX} \times{\variable WIDTH})$.
\item[{\variable xMASSINF}] 
 Lower limit of the Higgs ({\variable x=H}) or vector boson 
 ({\variable x=V1,V2}) mass range; used only when {\variable xGAMMAX} $<0$.
\item[{\variable xMASSSUP}] 
 Upper limit of the Higgs ({\variable x=H}) or vector boson 
 ({\variable x=V1,V2}) mass range; used only when {\variable xGAMMAX} $<0$.
\item[{\variable Vud}]
 CKM matrix elements, with {\variable u}={\variable U,C,T} and
 {\variable d}={\variable D,S,B}. Set {\variable VUD=VUS=VUB}=0
 to use values of PDG2003. {\variable Vud} is only used in single-$t$
 production.
\item[{\variable AEMRUN}]
 Set it to {\variable YES} to use running $\alpha_{em}$ in lepton pair and
 single vector boson production, set it to {\variable NO} to use 
 $\alpha_{em}=1/137.0359895$.
\item[{\variable IPROC}]
 Process number that identifies the hard subprocess: see table~\ref{tab:proc} 
 for valid entries.
\item[{\variable IVCODE}]
 Identifies the nature of the vector boson in associated Higgs production.
 It corresponds to variable {\variable IV} of table~\ref{tab:proc}.
\item[{\variable ILxCODE}]
 Identify the nature of the leptons emerging from vector boson decays\\
 ({\variable x} $=1,2$). They correspond to variables {\variable IL}$_1$ 
 and {\variable IL}$_2$ of table~\ref{tab:proc}.
\item[{\variable PARTn}]
 The type of the incoming particle \#{\variable n}, with {\variable n}=1,2. 
 \HW\ naming conventions are used ({\variable P, PBAR, N, NBAR}).
\item[{\variable PDFGROUP}]
 The name of the group fitting the parton densities used;
 the labeling conventions of PDFLIB are adopted. Unused when linked
 to LHAPDF.
\item[{\variable PDFSET}] 
 The number of the parton density set; according to PFDLIB conventions,
 the pair ({\variable PDFGROUP}, {\variable PDFSET}) identifies the 
 densities for a given particle type. When linked to LHAPDF, use 
 the numbering conventions of LHAGLUE~\cite{Whalley:2005nh}.
\item[{\variable LAMBDAFIVE}]
 The value of $\Lambda_{\sss QCD}$, for five flavours and in the 
 ${\overline {\rm MS}}$ scheme, used in the computation of NLO
 cross sections.
\item[{\variable LAMBDAHERW}]
 The value of $\Lambda_{\sss QCD}$ used in MC runs; this parameter has the 
 same meaning as $\Lambda_{\sss QCD}$ in \HW.
\item[{\variable SCHEMEOFPDF}] 
 The subtraction scheme in which the parton densities are defined.
\item[{\variable FPREFIX}] Our integration routine creates files with
 name beginning by the string {\variable FPREFIX}. These files are not 
 directly accessed by the user; for more details, see sect.~\ref{sec:evfile}.
\item[{\variable EVPREFIX}] 
 The name of the event file begins with this string; for 
 more details, see sect.~\ref{sec:evfile}.
\item[{\variable EXEPREFIX}] 
 The names of the \NLO\ and \MC\ executables begin with this string; this is
 useful in the case of simultaneous runs.
\item[{\variable NEVENTS}] 
 The number of events stored in the event file, eventually
 processed by \mbox{\HW\ .}
\item[{\variable WGTTYPE}]
 Valid entries are 0 and 1. When set to 0, the weights in the event file
 are $\pm 1$. When set to 1, they are $\pm w$, with $w$ a constant such
 that the sum of the weights gives the total NLO cross section.
 N.B.\ These weights are redefined by \HW\ at \MC\ run time according 
 to its own convention (see \HW\ manual).
\item[{\variable RNDEVSEED}] 
 The seed for the random number generation in the
 event generation step; must be changed in order to obtain
 statistically-equivalent but different event files.
\item[{\variable BASES}] 
 Controls the integration step; valid entries are {\variable ON} and 
 {\variable OFF}. At least one run with {\variable BASES=ON} must be 
 performed (see sect.~\ref{sec:evfile}).
\item[{\variable PDFLIBRARY}] 
 Valid entries are {\variable PDFLIB}, {\variable LHAPDF}, and 
 {\variable THISLIB}. In the former two case, PDFLIB or LHAPDF is used to 
 compute the parton densities, whereas in the latter case the densities are
 obtained from our self-contained faster package.
\item[{\variable HERPDF}] 
 If set to {\variable DEFAULT}, \HW\ uses its internal PDF set 
 (controlled by {\variable NSTRU}), regardless of the densities
 adopted at the NLO level. If set to {\variable EXTPDF}, \HW\ uses
 the same PDFs as the \NLO\ code (see sect.~\ref{sec:pdfs}).
\item[{\variable HWPATH}]
 The physical address of the directory where the user's
 preferred version of \HW\ is stored.
\item[{\variable SCRTCH}]
 The physical address of the directory where the user wants to store the
 data and event files. If left blank, these files are stored in the 
 running directory.
\item[{\variable HWUTI}]
This variables must be set equal to a list of object files,
needed by the analysis routines of the user (for example,
{\variable HWUTI=obj1.o obj2.o obj3.o} is a valid assignment).
\item[{\variable HERWIGVER}]
This variable must to be set equal to the name of the
object file corresponding to the version of \HW\ linked
to the package (for example, {\variable HERWIGVER=herwig65.o} is a
valid assignment).
\item[{\variable PDFPATH}]
 The physical address of the directory where the PDF grids are stored.
 Effective only if {\variable PDFLIBRARY=THISLIB}.
\item[{\variable LHAPATH}]
 Set this variable equal to the name of the directory where the local 
 version of LHAPDF is installed.
\item[{\variable LHAOFL}]
 Set {\variable LHAOFL=FREEZE} to freeze PDFs from LHAPDF at the boundaries,
 or equal to {\variable EXTRAPOLATE} otherwise. See LHAPDF manual for
 details.
\end{itemize}

\section*{Acknowledgement}
Many thanks to P.~Nason for contributions to the heavy quark code
and valuable discussions on all aspects of the $\MCatNLO$ project,
and to E.~Laenen and P.~Motylinski for the work done on single-$t$
production. We also thank V.~Drollinger and B.~Quayle for testing a 
preliminary version of the $W^+W^-$ code with spin correlations.

\section*{Appendices}
\appendix

\section{Version changes}

\subsection{From MC@NLO version 1.0 to version 2.0\label{app:newver}}
In this appendix we list the changes that occurred in the package
from version 1.0 to version 2.0.

$\bullet$~The Les Houches generic user process interface has been adopted.

$\bullet$~As a result, the convention for process codes has been changed:
MC@NLO process codes {\variable IPROC} are negative.

$\bullet$~The code {\code mcatnlo\_hwhvvj.f}, which was specific to
vector boson pair production in version 1.0, has been replaced by
{\code mcatnlo\_hwlhin.f}, which reads the event file according
to the Les Houches prescription, and works for all the production 
processes implemented.

$\bullet$~The {\code Makefile} need not be edited, since the variables
{\variable HERWIGVER} and {\variable HWUTI} have been moved to
{\code MCatNLO.inputs} (where they must be set by the user).

$\bullet$~A code {\code mcatnlo\_hbook.f} has been added to the list of
utility codes. It contains a simplified version (written by M.~Mangano)
of {\small HBOOK}, and it is only used by the sample analysis routines
{\code mcatnlo\_hwan{\em xxx}.f}. As such, the user will not need it
when linking to a self-contained analysis code.

We also remind the reader that the \HW\ version must be 
6.5 or higher since the  Les Houches interface is used.

\subsection{From MC@NLO version 2.0 to version 2.1\label{app:newvera}}
In this appendix we list the changes that occurred in the package
from version 2.0 to version 2.1.

$\bullet$~Higgs production has been added, which implies new process-specific 
files ({\code mcatnlo\_hgmain.f}, {\code mcatnlo\_hgxsec.f}, 
{\code hgscblks.h}, {\code mcatnlo\_hwanhgg.f}), and a modification to
{\code mcatnlo\_hwlhin.f}. 

$\bullet$~Post-1999 PDF sets have been added to the MC@NLO PDF library.

$\bullet$~Script variables have been added to {\code MCatNLO.inputs}. 
Most of them are only relevant to Higgs production, and don't affect processes
implemented in version 2.0. One of them ({\variable LAMBDAHERW}) may affect
all processes: in version 2.1, the variables {\variable LAMBDAFIVE} and
{\variable LAMBDAHERW} are used to set the value of $\Lambda_{\sss QCD}$ in
NLO and MC runs respectively, whereas in version 2.0 {\variable LAMBDAFIVE}
controlled both. The new setup is necessary since modern PDF sets have
$\Lambda_{\sss QCD}$ values which are too large to be supported by \HW.
(Recall that the effect of using {\variable LAMBDAHERW} different from
{\variable LAMBDAFIVE} is beyond NLO.)

$\bullet$~The new script variable {\variable PDFPATH} should be set equal 
to the name of the directory where the PDF grid files (which can be downloaded
from the MC@NLO web page) are stored. At run time, when executing {\code
runNLO}, or {\code runMC}, or {\code runMCatNLO}, logical links to these
files will be created in the running directory (in version 2.0, this
operation had to be performed by the user manually).

$\bullet$~Minor bugs corrected in {\code mcatnlo\_hbook.f} and sample 
analysis routines.

\subsection{From MC@NLO version 2.1 to version 2.2\label{app:newverb}}
In this appendix we list the changes that occurred in the package
from version 2.1 to version 2.2.

$\bullet$~Single vector boson production has been added, which implies 
new process-specific files ({\code mcatnlo\_sbmain.f}, 
{\code mcatnlo\_sbxsec.f}, {\code svbcblks.h}, {\code mcatnlo\_hwansvb.f}), 
and a modification to {\code mcatnlo\_hwlhin.f}. 

$\bullet$~The script variables {\variable WWIDTH} and {\variable ZWIDTH}
have been added to {\code MCatNLO.inputs}. These denote the physical widths 
of the $W$ and $Z^0$ bosons, used to generate the mass distributions of
the vector bosons according to the Breit--Wigner function, in the case
of single vector boson production (vector boson pair production is
still implemented only in the zero-width approximation).

\subsection{From MC@NLO version 2.2 to version 2.3\label{app:newverc}}
In this appendix we list the changes that occurred in the package
from version 2.2 to version 2.3.

$\bullet$~Lepton pair production has been added, which implies 
new process-specific files ({\code mcatnlo\_llmain.f}, 
{\code mcatnlo\_llxsec.f}, {\code llpcblks.h}, {\code mcatnlo\_hwanllp.f}), 
and modifications to {\code mcatnlo\_hwlhin.f} and 
{\code mcatnlo\_hwdriver.f}.

$\bullet$~The script variable {\variable AEMRUN} has been added, since
the computation of single vector boson and lepton pair cross sections is
performed in the $\MSbar$ scheme (the on-shell scheme was previously
used for single vector boson production).

$\bullet$~The script variables {\variable FRENMC} and {\variable FFACTMC}
have been eliminated. 

$\bullet$~The structure of pseudo-random number generation in heavy
flavour production has been changed, to avoid a correlation that
affected the azimuthal angle distribution for the products of the hard
partonic subprocesses.

$\bullet$~A few minor bugs have been corrected, which affected the rapidity
of the vector bosons in single vector boson production (a 2--3\% effect),
and the assignment of $\Lambda_{\sss QCD}$ for the LO and NLO PDF sets of
Alekhin.

\subsection{From MC@NLO version 2.3 to version 3.1\label{app:newverd}}
In this appendix we list the changes that occurred in the package
from version 2.3 to version 3.1.

$\bullet$~Associated Higgs production has been added, which implies 
new process-specific files ({\code mcatnlo\_vhmain.f}, 
{\code mcatnlo\_vhxsec.f}, {\code vhgcblks.h}, {\code mcatnlo\_hwanvhg.f}), 
and modifications to {\code mcatnlo\_hwlhin.f} and 
{\code mcatnlo\_hwdriver.f}.

$\bullet$~Spin correlations in $W^+W^-$ production and leptonic decay
have been added;
the relevant codes ({\code mcatnlo\_vpmain.f}, {\code mcatnlo\_vhxsec.f})
have been modified; the sample analysis routines ({\code mcatnlo\_hwanvbp.f})
have also been changed. Tree-level matrix elements have been computed with
MadGraph/MadEvent~\cite{Stelzer:1994ta,Maltoni:2002qb}, which uses 
HELAS~\cite{Murayama:1992gi}; the relevant routines and common blocks 
are included in {\code mcatnlo\_helas2.f} and {\code MEcoupl.inc}.

$\bullet$~The format of the event file has changed in several respects,
the most relevant of which is that the four-momenta are now given as
$(p_x,p_y,p_z,m)$ (up to version 2.3 we had $(p_x,p_y,p_z,E)$). Event
files generated with version 2.3 or lower {\em must not be used} with
version 3.1 or higher (the code will prevent the user from doing so).

$\bullet$~The script variables {\variable GAMMAX}, {\variable MASSINF},
and {\variable MASSSUP} have been replaced with {\variable xGAMMAX}, 
{\variable xMASSINF} and {\variable xMASSSUP}, with {\variable x=H,V1,V2}.

$\bullet$~New script variables {\variable IVCODE}, {\variable IL1CODE},
and {\variable IL2CODE} have been introduced.

$\bullet$~Minor changes have been made to the routines that put the partons
on the \HW\ mass shell for lepton pair, heavy quark, and vector boson pair
production; effects are beyond the fourth digit.

$\bullet$~The default electroweak parameters have been changed for
vector boson pair production, in order to make them consistent with those
used in other processes. The cross sections are generally smaller
in version 3.1 wrt previous versions, the dominant effect being the 
value of $\sinthW$: we have now $\sinsqthW=0.2311$, in lower
versions $\sinsqthW=1-m_W^2/m_Z^2$. The cross sections
are inversely proportional to $\sinfthW$.

\subsection{From MC@NLO version 3.1 to version 3.2\label{app:newvere}}
In this appendix we list the changes that occurred in the package
from version 3.1 to version 3.2.

$\bullet$~Single-$t$ production has been added, which implies 
new process-specific files ({\code mcatnlo\_stmain.f}, 
{\code mcatnlo\_stxsec.f}, {\code stpcblks.h}, {\code mcatnlo\_hwanstp.f}), 
and modifications to {\code mcatnlo\_hwlhin.f} and 
{\code mcatnlo\_hwdriver.f}.

$\bullet$~LHAPDF library is now supported, which implies modifications to
all {\code *main.f} files, and two new utility codes,
{\code mcatnlo\_lhauti.f} and {\code mcatnlo\_mlmtolha.f}.

$\bullet$~New script variables {\variable Vud}, {\variable LHAPATH},
and {\variable LHAOFL} have been introduced.

$\bullet$~A bug affecting Higgs production has been fixed, which implies
a modification to {\code mcatnlo\_hgxsec.f}. Cross sections change with
respect to version 3.1 {\em only if} {\variable FFACT}$\ne 1$ (by 
${\cal O}(1\%)$ in the range $1/2\le$ {\variable FFACT} $\le 2$).

\section{Running the package without the shell scripts\label{app:instr}}
In this appendix, we describe the actions that the user needs to 
take in order to run the package without using the shell scripts,
and the {\variable Makefile}. Examples are given for vector boson
pair production, but only trivial modifications are necessary in
order to treat other production processes.

\subsection{Creating the executables\label{app:exe}}
An $\MCatNLO$ run requires the creation of two executables, for the \NLO\
and \MC\ codes respectively. The files to link depend on whether one
uses PDFLIB, LHAPDF, or the PDF library provided with this package; 
we list them below:
\begin{itemize}
\item {\bf NLO with private PDFs:}
{\code mcatnlo\_vbmain.o mcatnlo\_vbxsec.o mcatnlo\_helas2.o 
mcatnlo\_date.o mcatnlo\_int.o mcatnlo\_uxdate.o mcatnlo\_uti.o 
mcatnlo\_str.o mcatnlo\_pdftomlm.o mcatnlo\_libofpdf.o dummies.o SYSFILE}
\item {\bf NLO with PDFLIB:}
{\code mcatnlo\_vbmain.o mcatnlo\_vbxsec.o mcatnlo\_helas2.o 
mcatnlo\_date.o mcatnlo\_int.o mcatnlo\_uxdate.o mcatnlo\_uti.o 
mcatnlo\_str.o mcatnlo\_mlmtopdf.o dummies.o}
{\variable SYSFILE CERNLIB}
\item {\bf NLO with LHAPDF:}
{\code mcatnlo\_vbmain.o mcatnlo\_vbxsec.o mcatnlo\_helas2.o 
mcatnlo\_date.o mcatnlo\_int.o mcatnlo\_uxdate.o mcatnlo\_lhauti.o 
mcatnlo\_str.o mcatnlo\_mlmtolha.o dummies.o}
{\variable SYSFILE LHAPDF}
\item {\bf MC with private PDFs:}
{\code mcatnlo\_hwdriver.o mcatnlo\_hwlhin.o mcatnlo\_hwanvbp.o 
mcatnlo\_hbook.o mcatnlo\_str.o mcatnlo\_pdftomlm.o mcatnlo\_libofpdf.o 
dummies.o} {\variable HWUTI HERWIGVER}
\item {\bf MC with PDFLIB:}
{\code mcatnlo\_hwdriver.o mcatnlo\_hwlhin.o mcatnlo\_hwanvbp.o 
mcatnlo\_hbook.o mcatnlo\_str.o mcatnlo\_mlmtopdf.o dummies.o}
{\variable HWUTI HERWIGVER CERNLIB}
\item {\bf MC with LHAPDF:}
{\code mcatnlo\_hwdriver.o mcatnlo\_hwlhin.o mcatnlo\_hwanvbp.o 
mcatnlo\_hbook.o mcatnlo\_str.o mcatnlo\_mlmtolha.o dummies.o}
{\variable HWUTI HERWIGVER LHAPDF}
\end{itemize}
The process-specific codes {\code mcatnlo\_vbmain.o} and
{\code mcatnlo\_vbxsec.o} (for the \NLO\ executable) and
{\code mcatnlo\_hwanvbp.o} (the \HW\ analysis routines in the
\MC\ executable) need to be replaced by their analogues for 
other production processes, which can be easily read from the
list given in sect.~\ref{sec:packfile}.

The variable {\variable SYSFILE} must be set either equal to {\code alpha.o},
or to {\code linux.o}, or to {\code sun.o}, according to the architecture 
of the machine on which the run is performed. For any other architecture,
the user should provide a file corresponding to {\code alpha.f} etc.,
which he/she will easily obtain by modifying {\code alpha.f}. The 
variables {\variable HWUTI} and {\variable HERWIGVER} have been described
in sect.~\ref{sec:scrvar}. In order to create the object files eventually 
linked, static compilation is always recommended (for example, 
{\code g77 -Wall -fno-automatic} on Linux).

\subsection{The input files\label{app:input}}
In this appendix, we describe the inputs to be given to the \NLO\ and 
\MC\ executables in the case of vector boson pair production. The case
of other production processes is completely analogous.
When the shell scripts are used to run the $\MCatNLO$,
two files are created, {\variable FPREFIXNLOinput} and 
{\variable FPREFIXMCinput}, which are read by the \NLO\ and \MC\ executable
respectively. We start by considering the inputs for the \NLO\
executable, presented in table~\ref{tab:NLOi}.
\begin{table}[htb]
\begin{center}
\begin{tabular}{ll}
\hline
 '{\variable FPREFIX}'                       & ! prefix for BASES files\\
 '{\variable EVPREFIX}'                      & ! prefix for event files\\
  {\variable ECM FFACT FREN FFACTMC FRENMC}  & ! energy, scalefactors\\
  {\variable IPROC}                        & ! -2850/60/70/80=WW/ZZ/ZW+/ZW-\\
  {\variable WMASS ZMASS}                    & ! M\_W, M\_Z\\
  {\variable UMASS DMASS SMASS CMASS BMASS GMASS} & ! quark and gluon masses\\
 '{\variable PART1}'  '{\variable PART2}'    & ! hadron types\\
 '{\variable PDFGROUP}'   {\variable PDFSET} & ! PDF group and id number\\
  {\variable LAMBDAFIVE}                     & ! Lambda\_5, $<$0 for default\\
 '{\variable SCHEMEOFPDF}'                   & ! scheme\\
  {\variable NEVENTS}                        & ! number of events\\
  {\variable WGTTYPE}                 & ! 0 =$>$ wgt=+1/-1, 1 =$>$ wgt=+w/-w\\
  {\variable RNDEVSEED}                      & ! seed for rnd numbers\\
  {\variable zi}                             & ! zi\\
  {\variable nitn$_1$ nitn$_2$}              & ! itmx1,itmx2\\
\hline\\
\end{tabular}
\end{center}
\caption{\label{tab:NLOi}
Sample input file for the \NLO\ code (for vector boson pair production). 
{\variable FPREFIX} and {\variable EVPREFIX} must be understood with 
{\variable SCRTCH} in front (see sect.~\ref{sec:scrvar}).
}
\end{table}
The variables whose name is in uppercase characters have been described 
in sect.~\ref{sec:scrvar}. The other variables are assigned by the shell
script. Their default values are given in table~\ref{tab:defNLO}.
\begin{table}[htb]
\begin{center}
\begin{tabular}{ll}
\hline
Variable & Default value\\
\hline
{\variable zi}          & 0.2\\
{\variable nitn$_i$}    & 10/0 ({\variable BASES=ON/OFF})\\
\hline\\
\end{tabular}
\end{center}
\caption{\label{tab:defNLO}
Default values for script-generated variables in {\code FPREFIXNLOinput}.
}
\end{table}
Users who run the package without the script should use the values
given in table~\ref{tab:defNLO}. The variable {\variable zi} controls,
to a certain extent, the number of negative-weight events generated 
by the $\MCatNLO$ (see ref.~\cite{Frixione:2002ik}). Therefore, the user
may want to tune this parameter in order to reduce as much as possible
the number of negative-weight events. We stress that the \MC\ code will
not change this number; thus, the tuning can (and must) be done only 
by running the \NLO\ code. The variables {\variable nitn$_i$} control
the integration step (see sect.~\ref{sec:evfile}), which can be
skipped by setting {\variable nitn$_i=0$}. If one needs to perform the
integration step, we suggest setting these variables as indicated in
table~\ref{tab:defNLO}. 

\begin{table}[htb]
\begin{center}
\begin{tabular}{ll}
\hline
 '{\variable EVPREFIX.events}'               & ! event file\\
  {\variable NEVENTS}                        & ! number of events\\
  {\variable pdftype}                      & ! 0-$>$Herwig PDFs, 1 otherwise\\
 '{\variable PART1}'  '{\variable PART2}'    & ! hadron types\\
  {\variable beammom beammom}                & ! beam momenta\\
  {\variable IPROC}                         & ! --2850/60/70/80=WW/ZZ/ZW+/ZW-\\
 '{\variable PDFGROUP}'                      & ! PDF group (1)\\
  {\variable PDFSET}                         & ! PDF id number (1)\\
 '{\variable PDFGROUP}'                      & ! PDF group (2)\\
  {\variable PDFSET}                         & ! PDF id number (2)\\
  {\variable LAMBDAHERW}                     & ! Lambda\_5, $<$0 for default\\
  {\variable WMASS WMASS ZMASS}              & ! M\_W+, M\_W-, M\_Z\\
  {\variable UMASS DMASS SMASS CMASS BMASS GMASS} & ! quark and gluon masses\\
\hline\\
\end{tabular}
\end{center}
\caption{\label{tab:MCi}
Sample input file for the \MC\ code (for vector boson pair production), 
resulting from setting {\variable HERPDF=EXTPDF}, which implies 
{\variable pdftype=1}. 
Setting {\variable HERPDF=DEFAULT} results in an analogous file, with
{\variable pdftype=0}, and without the lines concerning
{\variable PDFGROUP} and {\variable PDFSET}. {\variable EVPREFIX} 
must be understood with {\variable SCRTCH} in front 
(see sect.~\ref{sec:scrvar}). The negative sign of {\variable IPROC}
tells \HW\ to use Les Houches interface routines.
}
\end{table}
We now turn to the inputs for the \MC\ executable, presented
in table~\ref{tab:MCi}. 
The variables whose names are in uppercase characters have been described 
in sect.~\ref{sec:scrvar}. The other variables are assigned by the shell
script. Their default values are given in table~\ref{tab:defMC}.
\begin{table}[htb]
\begin{center}
\begin{tabular}{ll}
\hline
Variable & Default value\\
\hline
{\variable esctype}         & 0\\
{\variable pdftype}         & 0/1 ({\variable HERPDF=DEFAULT/EXTPDF})\\
{\variable beammom}         & {\variable EMC}/2\\
\hline\\
\end{tabular}
\end{center}
\caption{\label{tab:defMC}
Default values for script-generated variables in {\code MCinput}.
}
\end{table}
The user can freely change the values of {\variable esctype} and
{\variable pdftype}; on the other hand, the value of {\variable beammom}
must always be equal to half of the hadronic CM energy.

When LHAPDF is linked, the value of {\variable PDFSET} is sufficient
to identify the parton density set. In such a case, {\variable PDFGROUP}
must be set in input equal to {\variable LHAPDF} if the user wants
to freeze the PDFs at the boundaries (defined as the ranges in which
the fits have been performed). If one chooses to extrapolate the PDFs
across the boundaries, one should set {\variable PDFGROUP=LHAEXT}
in input.

In the case of $\gamma/Z$, $W^\pm$, Higgs or heavy quark production, the 
\MC\ executable can be run with the corresponding positive input process 
codes {\variable IPROC} = 1350, 1399, 1499, 1600+ID, 1705, 1706,
2000--2008, 2600+ID or 2700+ID,
to generate a standard \HW\ run for comparison purposes\footnote{For
vector boson pair production, for historical reasons, the different
process codes 2800--2825 must be used.}.  Then the input
event file will not be read: instead, parton configurations will be
generated by \HW\ according to the LO matrix elements.


\end{document}